    \documentclass[aps,prb,groupedaddress,twocolumn,amsmath,floatfix]{revtex4}
   \usepackage{graphicx}
   \usepackage{longtable}
\newcommand\beq{\begin{equation}}
\newcommand\eeq{\end{equation}}
\newcommand\beqa{\begin{eqnarray}}
\newcommand\eeqa{\end{eqnarray}}
\newcommand\nn{\nonumber\\}

\begin{document}

\title{Pair correlation function of short-ranged square-well fluids}

\author{J. Largo}
\email[]{largoju@unican.es}
\author{J. R. Solana}
\email[]{solanajr@unican.es}
\thanks{Author to whom correspondence should be addressed}
\affiliation{Departamento de F\'{\i}sica Aplicada, Universidad de
Cantabria, E-39005 Santander, Spain}
\author{S. B.Yuste}
\email[]{santos@unex.es}
\homepage[]{http://www.unex.es/fisteor/santos/sby}
\author{A. Santos}
\email[]{andres@unex.es}
\homepage[]{http://www.unex.es/fisteor/andres/}
\affiliation{Departamento de F\'{\i}sica, Universidad de
Extremadura, E-06071 Badajoz, Spain}


\date{\today}

\begin{abstract}
 We have performed extensive Monte Carlo simulations in the canonical
(NVT) ensemble of the pair correlation function for square-well
fluids with well widths $\lambda-1$ ranging from $0.1$ to $1.0$, in
units of the diameter $\sigma$ of the particles. For each one of
these widths, {several} densities $\rho$ and temperatures $T$ in the
ranges $0.1\leq\rho\sigma^3\leq 0.8$ and $T_c(\lambda)\lesssim
T\lesssim 3T_c(\lambda)$, where $T_c(\lambda)$ is the critical
temperature, have been considered. The simulation data are used to
examine the performance of two analytical theories in predicting the
structure of these fluids: the perturbation theory proposed by Tang
and Lu [{Y. Tang and B.~C.-Y. Lu},  { J. Chem. Phys.} {\bf 100},
3079, 6665 (1994)] and the non-perturbative model proposed by two of
us [{S. B. Yuste and A. Santos},  { J. Chem. Phys.} {\bf 101}, 2355
(1994)]. It is observed that both theories complement each other, as
the latter theory works well for short ranges and/or moderate
densities, while the former theory does for long ranges and high
densities.
\end{abstract}


\maketitle

\section{Introduction\label{sec1}}

The study of the thermodynamic and structural properties of
square-well (SW) fluids has been the subject of interest for many
years because of their simplicity and their resemblance with real
fluids with spherically symmetrical potentials, among other reasons.
Therefore, at present there are available a considerable number of
theories for this kind of fluids. Among them, particularly simple
and fruitful are perturbation theories for the thermodynamic
properties.\cite{BH:67,SHB:70,SHB:71,HBS:76,HSS:80,BH:76,CS:94,BG:99,LS:03-a,LSAS:03}
If  one is interested in structural properties, one can resort to
integral equations theories based on the Ornstein--Zernike equation,
 for
which we have a number of possible
choices.\cite{T:73a,T:73b,T:74,HMF:76,SHT:77,SH:78,JKLFF:81,KH:82,STMM:93,GVRM:95,BWWA:96,C:96,LKLLW:99,RK:02}
The latter group of theories have in general the drawback of being
non-analytical, so one {has to deal with} them by numerical methods.
However, in some cases it has been possible to obtain analytical
expressions for the structural
properties\cite{N:77,SS:77,TL:94-a,TL:94-b,BYS:94,AS:01} inspired,
at least indirectly, in integral equation theories.

In parallel with the theoretical developments, a lot of research has
been devoted to obtaining the thermodynamic and structural
properties of SW fluids by means of computer
simulations.\cite{HSS:80,LSAS:03,HMF:76,R:65,AYM:72,RT:75,YA:80,CMC:87,LLA:90,BAR:91,HA:92,VMRJM:92,
M:97,BV:98-a,OP:99,EH:99,OFP:01,LMKSR:99,RAERJL:02,SKE:03,LS:03-b}

Most of that research has focused on SW fluid with intermediate
ranges of the potential, because they more closely mimic real simple
fluids, whereas relatively little attention has been paid to SW
fluids with short ranges. On the other hand,  recently there has
been a renewal in the interest in short-ranged SW fluids as models
of colloidal
suspensions\cite{LKLLW:99,HSKGKQ:84,ZFDST:01,DFFGSSTVZ:01,ZFDBST:03}
and phase separation of protein solutions.\cite{LAB:96}

In the present paper we have carried out Monte Carlo simulations of
the pair correlation function or radial distribution function
(r.d.f.) $g(r)$ of SW fluids with short, intermediate, and long
ranges  for temperatures above the critical ones and for a wide
range of densities. These data are used to test the performance of
two analytical theories, one perturbative\cite{TL:94-a,TL:94-b} and
the other one non-perturbative.\cite{BYS:94} As  we will see, both
theories complement each other: the perturbative theory is generally
preferable for long ranges, while the non-perturbative theory is
better for short ranges.

The plan of the paper is as follows. The two theories are
introduced in the next Section, some details being relegated to
Appendices \ref{appA} and \ref{appB}. The Monte Carlo method we have
employed is succinctly described in Section \ref{sec3}. The
theoretical results are compared with the simulation data and
discussed in Section \ref{sec4}. The main conclusions of the paper
are summarized in Section \ref{sec5}.

\section{Analytical theories for the pair correlation function of square-well fluids}

\label{sec2}
 For fluids with a square-well potential of the form
\begin{equation}
u\left( r \right)=\begin{cases} \infty & \mbox{if}\quad r\le
\sigma,\\
  -\epsilon & \mbox{if}\quad \sigma <r\le \lambda \sigma ,\\
   0& \mbox{if}\quad r>\lambda \sigma ,
   \end{cases}
  \label{eq:u(r)SW}
\end{equation}
where $\lambda$ is the potential range in units of the particle
diameter $\sigma$ and $\epsilon$ is the potential depth, several
approaches have been devised to derive analytical expressions for
the structural properties. In this paper we will focus on two
theories, both having in common that analytical expressions for the
r.d.f. in Laplace space are provided.

The first one of those theories is due to Tang and Lu
(TL),\cite{TL:94-a,TL:94-b} who combined perturbation theory with
the mean spherical approximation (MSA) to derive an analytical
expression for the first-order r.d.f. $g_1(r)$ in the expansion in
power series of the inverse of the reduced temperature
$T^{*}=kT/\epsilon$. Taking for the zeroth-order term $g_0(r)$ the
Percus--Yevick (PY) solution,\cite{W:63,T:63} the resulting
truncated series for the r.d.f. of the SW fluid is
\begin{equation}
g(x) = g_{0}(x) + g_{1}(x) \frac{1}{T^{*}},
\label{eq:g(x)-per}
\end{equation}
 where $x=r/\sigma$. The expressions for the Laplace transforms of
$xg_0(x)$ and $xg_1(x)$ are given in Appendix \ref{appA}. The TL
theory is expected to be accurate for moderate to large potential
widths since the series in powers of $1/T^*$ converges slowly for
short-ranged SW potentials.\cite{LS:03}

On the opposite situation, that is,  for SW potential with ranges
$\lambda$ close to $1$, a procedure has been proposed\cite{SR:94} to
determine the structure of a SW fluid from that of an equivalent
fluid of sticky hard spheres, using for the latter Baxter's
analytical solution of the PY equation.\cite{B:68} To this end, the
parameters of the equivalent fluid are determined from the condition
that the second virial coefficients of the two fluids must be equal.
This approximation provides good results for the structure factor of
SW fluids with  $\lambda \leq 1.2$, at least for moderate to low
densities, but is not appropriate for obtaining the r.d.f., {so that
it will not be considered in this paper.}

As a second theory, we will consider {the one developed by Yuste and
Santos,\cite{BYS:94} which provides an alternative analytical
expression for short-ranged SW fluids  and reduces to Baxter's
solution in the sticky hard-sphere limit.}  We will refer to this
theory as the Yuste--Santos (YS) model and it will be presented next
with some detail.

The starting point in the YS model is the expression of the Laplace
transform $G(t)$ of $xg(x)$ in the form
\begin{equation}
G\left( t \right)=t\frac{F\left( t \right)e^{-t}}{1+12\eta F\left( t
\right)e^{-t}}=\sum\limits_{n=1}^\infty  {\left( {-12\eta }
\right)^{n-1}t\left[ {F\left( t \right)} \right]^ne^{-nt}},
\label{eq:G(t)}
\end{equation}
where $\eta=\frac{\pi}{6}\rho\sigma^3$ is the packing fraction,
$\rho$ being the number density, and $F(t)$ is an auxiliary function
given by\cite{BYS:94,AS:01}
\begin{equation}
F(t)=-\frac{1}{12\eta}\frac{1+A+K_1 t-\left(A+K_2 t\right)e^{-(\lambda-1)t}}{1+S_1 t+S_2 t^2+S_3 t^3}.
\label{eq:F(t)}
\end{equation}
The coefficients $K_1$, $K_2$, $S_1$, $S_2$, and $S_3$ are
determined from consistency conditions as functions of $\eta$,
$T^*$, $\lambda$, and $A$ (see Appendix \ref{appB}). To close the
model, the parameter $A$ is further fixed at its zero density value
$A=e^{1/T^*}-1$
 for the sake of simplicity.\cite{BYS:94,AS:01} Expression
(\ref{eq:F(t)}) reduces to the exact solutions of the PY equation in
the limit of hard spheres ($\lambda\to 1$ or
$T^*\to\infty$),\cite{W:63,T:63} as well as in the limit of sticky
hard spheres ($\lambda\to 1$ and $T^*\to 0$ with $T^*\sim
-1/\ln(\lambda-1)$).\cite{B:68} Therefore, the approximation
(\ref{eq:F(t)}) can be considered  as an extension to finite widths
of Baxter's solution of the PY equation for sticky hard spheres.

The inverse Laplace transform of Eq.\ (\ref{eq:G(t)}) allows us to
obtain the r.d.f. in the form
\begin{equation}
g\left( x \right)=x^{-1}\sum\limits_{n=1}^\infty  {\left( {-12\eta }
\right)^{n-1}f_n\left( {x-n} \right)\Theta \left( {x-n} \right)},
\label{eq:g(r)-SW}
\end{equation}
where the function $f_{n}(x)$ is the inverse Laplace transform of
$t[F(t)]^{n}$ and $\Theta (x)$ is  Heaviside's step function. Note
that, to determine the r.d.f. for $x < n+1$ only the first $n$ terms
in the summation (\ref{eq:g(r)-SW}) are needed. In the analysis of
Section \ref{sec4}, we will consider reduced distances $x< 3$, so
that only the functions $f_{1}$ and $f_{2}$ will be needed.
They are given in Appendix \ref{appB}.

\section{Monte Carlo simulations\label{sec3}}

We have performed {NVT} Monte Carlo (MC) simulations of the r.d.f.
of SW fluids with ranges $\lambda = 1.05$ and  $\lambda =1.1$--$2.0$
(with a step $\Delta\lambda=0.1$) for (reduced) number densities
$\rho^*\equiv\rho\sigma^3=0.1$--$0.8$ (with a step
$\Delta\rho^*=0.1$)  and several temperatures in the supercritical
region.  To this end, a system consisting of $512$ particles was
considered. The particles were initially placed in  a regular
configuration in a cubic volume with periodic boundary conditions,
with fixed  temperature and density. After equilibration, the r.d.f.
was determined from measurements performed over $5 \times 10^{4}$
cycles, each of them consisting of an attempted move per particle.
Results for the contact values $g(1^{+})$ of the r.d.f., as well as
for their values $g(\lambda^{-})$ and $g(\lambda^{+})$ at both sides
of the potential range, were obtained from extrapolation and are
reported in Table \ref{Table g(1+)-g(lambda+-)}.\cite{note} {}From
these values, the compressibility factor $Z=pV/NkT$ can be obtained
from the virial theorem for the SW fluid as
\begin{equation}
Z=1+\frac{2}{3}\pi\rho^{*}\left\{g(1^+)-\lambda^{3}\left[g(\lambda^{-})-g(\lambda^{+})\right]\right\}.
\label{VT-SW}
\end{equation}
Values of $Z$ thus obtained were reported elsewhere,\cite{LS:03-b}
except for the range $\lambda=1.05$.
\begingroup
\squeezetable
\begin{longtable}{lccccccccc}
\caption{MC simulation data of $g(1^{+})$, $g(\lambda^{-})$, and
$g(\lambda^{+})$.
}
\\
\hline\hline
$\rho^{*}$ & $0.10$ & $0.20$ & $0.30$ & $0.40$ & $0.50$ & $0.60$ & $0.70$ & $0.80$ & $0.90$\\
\hline
\endfirsthead
\caption{Continued.}
\\
\hline\hline
$\rho^{*}$ & $0.10$ & $0.20$ & $0.30$ & $0.40$ & $0.50$ & $0.60$ & $0.70$ & $0.80$ & $0.90$\\
\hline
\endhead
\endfoot
\hline\hline
\endlastfoot
\\
\multicolumn{10}{c} {$\lambda=1.05$}\\
 $T^{*}=0.5$\\
\cline{1-1}
$g(1^{+})$ & $7.304$ & $7.323$ & $7.343$ & $7.501$ & $7.662$ & $7.771$ & $8.081$ & $8.524$ & $9.078$\\
$g(\lambda^{-})$ & $7.278$ & $7.250$ & $7.210$ & $7.317$ & $7.374$ & $7.400$ & $7.587$ & $7.832$ & $8.154$\\
$g(\lambda^{+})$ & $0.987$ & $0.980$ & $0.979$ & $0.988$ & $0.999$ & $0.999$ & $1.030$ & $1.059$ & $1.105$\\
\\ $T^{*}=0.7$\\
\cline{1-1}
$g(1^{+})$ & $4.382$ & $4.609$ & $4.899$ & $5.194$ & $5.557$ & $6.009$ & $6.505$ & $7.223$ & $8.065$\\
$g(\lambda^{-})$ & $4.321$ & $4.562$ & $4.765$ & $4.986$ & $5.237$ & $5.539$ & $5.842$ & $6.263$ & $6.657$\\
$g(\lambda^{+})$ & $1.034$ & $1.081$ & $1.142$ & $1.195$ & $1.257$ & $1.325$ & $1.401$ & $1.504$ & $1.595$\\
\\ $T^{*}=1.0$\\
\cline{1-1}
$g(1^{+})$ & $2.964$ & $3.215$ & $3.552$ & $3.891$ & $4.336$ & $4.833$ & $5.457$ & $6.258$ & $7.245$\\
$g(\lambda^{-})$ & $2.923$ & $3.153$ & $3.417$ & $3.679$ & $4.009$ & $4.356$ & $4.759$ & $5.197$ & $5.666$\\
$g(\lambda^{+})$ & $1.071$ & $1.162$ & $1.255$ & $1.355$ & $1.477$ & $1.599$ & $1.746$ & $1.912$ & $2.085$\\
\hline
\\ \multicolumn{10}{c} {$\lambda=1.10$}\\
 $T^{*}=0.5$\\
\cline{1-1}
$g(1^{+})$ & $7.254$ & $7.080$ & $6.628$ & $6.461$ & $6.088$ & $5.836$ & $5.667$ & $5.620$ & $5.810$\\
$g(\lambda^{-})$ & $7.179$ & $6.919$ & $6.463$ & $6.219$ & $5.816$ & $5.519$ & $5.307$ & $5.180$ & $5.162$\\
$g(\lambda^{+})$ & $0.973$ & $0.931$ & $0.875$ & $0.840$ & $0.787$ & $0.746$ & $0.718$ & $0.701$ & $0.698$\\
\\ $T^{*}=0.7$\\
\cline{1-1}
$g(1^{+})$ & $4.135$ & $4.153$ & $4.174$ & $4.222$ & $4.334$ & $4.529$ & $4.746$ & $5.064$ & $5.604$\\
$g(\lambda^{-})$ & $4.097$ & $4.064$ & $4.028$ & $4.007$ & $4.028$ & $4.096$ & $4.135$ & $4.202$ & $4.284$\\
$g(\lambda^{+})$ & $0.979$ & $0.970$ & $0.964$ & $0.961$ & $0.966$ & $0.980$ & $0.988$ & $1.007$ & $1.027$\\
\\ $T^{*}=1.0$\\
\cline{1-1}
$g(1^{+})$ & $2.829$ & $2.963$ & $3.134$ & $3.320$ & $3.565$ & $3.868$ & $4.263$ & $4.778$ & $5.481$\\
$g(\lambda^{-})$ & $2.790$ & $2.862$ & $2.974$ & $3.068$ & $3.186$ & $3.316$ & $3.463$ & $3.589$ & $3.713$\\
$g(\lambda^{+})$ & $1.026$ & $1.058$ & $1.090$ & $1.128$ & $1.172$ & $1.220$ & $1.273$ & $1.316$ & $1.362$\\
\hline
\\ \multicolumn{10}{c} {$\lambda=1.20$}\\
 $T^{*}=0.7$\\
\cline{1-1}
$g(1^{+})$ & $4.126$ & $4.007$ & $3.805$ & $3.562$ & $3.449$ & $3.297$ & $3.281$ & $3.457$ & $3.921$\\
$g(\lambda^{-})$ & $4.030$ & $3.822$ & $3.555$ & $3.299$ & $3.122$ & $2.924$ & $2.810$ & $2.738$ & $2.612$\\
$g(\lambda^{+})$ & $0.968$ & $0.914$ & $0.850$ & $0.792$ & $0.748$ & $0.700$ & $0.675$ & $0.657$ & $0.628$\\
\\ $T^{*}=1.0$\\
\cline{1-1}
$g(1^{+})$ & $2.671$ & $2.662$ & $2.666$ & $2.721$ & $2.799$ & $2.941$ & $3.173$ & $3.564$ & $4.232$\\
$g(\lambda^{-})$ & $2.628$ & $2.551$ & $2.489$ & $2.447$ & $2.415$ & $2.388$ & $2.372$ & $2.326$ & $2.208$\\
$g(\lambda^{+})$ & $0.968$ & $0.942$ & $0.918$ & $0.900$ & $0.888$ & $0.880$ & $0.871$ & $0.856$ & $0.812$\\
\\ $T^{*}=1.5$\\
\cline{1-1}
$g(1^{+})$ & $2.016$ & $2.092$ & $2.212$ & $2.355$ & $2.541$ & $2.807$ & $3.168$ & $3.692$ & $4.513$\\
$g(\lambda^{-})$ & $1.956$ & $1.973$ & $1.991$ & $2.019$ & $2.042$ & $2.059$ & $2.063$ & $2.028$ & $1.917$\\
$g(\lambda^{+})$ & $1.004$ & $1.012$ & $1.021$ & $1.036$ & $1.050$ & $1.057$ & $1.059$ & $1.039$ & $0.982$\\
\hline
\\ \multicolumn{10}{c} {$\lambda=1.30$}\\
 $T^{*}=1.0$\\
\cline{1-1}
$g(1^{+})$ & $2.660$ & $2.592$ & $2.485$ & $2.435$ & $2.414$ & $2.492$ & $2.703$ & $3.136$ & $4.051$\\
$g(\lambda^{-})$ & $2.584$ & $2.433$ & $2.259$ & $2.138$ & $2.033$ & $1.948$ & $1.865$ & $1.737$ & $1.487$\\
$g(\lambda^{+})$ & $0.952$ & $0.896$ & $0.830$ & $0.786$ & $0.748$ & $0.716$ & $0.685$ & $0.639$ & $0.547$\\
\\ $T^{*}=1.5$\\
\cline{1-1}
$g(1^{+})$ & $1.957$ & $1.982$ & $2.033$ & $2.105$ & $2.245$ & $2.476$ & $2.829$ & $3.409$ & $4.390$\\
$g(\lambda^{-})$ & $1.872$ & $1.834$ & $1.792$ & $1.754$ & $1.720$ & $1.680$ & $1.611$ & $1.486$ & $1.272$\\
$g(\lambda^{+})$ & $0.966$ & $0.940$ & $0.919$ & $0.901$ & $0.883$ & $0.863$ & $0.828$ & $0.763$ & $0.652$\\
\\ $T^{*}=2.0$\\
\cline{1-1}
$g(1^{+})$ & $1.711$ & $1.770$ & $1.880$ & $2.012$ & $2.214$ & $2.497$ & $2.916$ & $3.558$ & $4.537$\\
$g(\lambda^{-})$ & $1.622$ & $1.618$ & $1.611$ & $1.606$ & $1.582$ & $1.555$ & $1.489$ & $1.368$ & $1.176$\\
$g(\lambda^{+})$ & $0.990$ & $0.981$ & $0.979$ & $0.972$ & $0.962$ & $0.944$ & $0.903$ & $0.829$ & $0.710$\\
\hline
\\ \multicolumn{10}{c} {$\lambda=1.40$}\\
 $T^{*}=1.0$\\
\cline{1-1}
$g(1^{+})$ & $2.899$ & $2.840$ & $2.640$ & $2.398$ & $2.246$ & $2.245$ & $2.515$ & $3.162$ & $4.351$\\
$g(\lambda^{-})$ & $2.679$ & $2.488$ & $2.227$ & $1.966$ & $1.783$ & $1.669$ & $1.544$ & $1.332$ & $1.055$\\
$g(\lambda^{+})$ & $0.986$ & $0.918$ & $0.819$ & $0.725$ & $0.657$ &
$0.616$ & $0.568$ & $0.491$ & $0.389$\\
\\ $T^{*}=1.5$\\
\cline{1-1}
$g(1^{+})$ & $1.928$ & $1.916$ & $1.917$ & $1.977$ & $2.075$ & $2.307$ & $2.716$ & $3.442$ & $4.571$\\
$g(\lambda^{-})$ & $1.836$ & $1.742$ & $1.656$ & $1.583$ & $1.518$ & $1.441$ & $1.328$ & $1.149$ & $0.932$\\
$g(\lambda^{+})$ & $0.945$ & $0.894$ & $0.849$ & $0.811$ & $0.778$ & $0.740$ & $0.681$ & $0.589$ & $0.478$\\
\\ $T^{*}=2.0$\\
\cline{1-1}
$g(1^{+})$ & $1.665$ & $1.705$ & $1.788$ & $1.889$ & $2.070$ & $2.362$ & $2.834$ & $3.584$ & $4.719$\\
$g(\lambda^{-})$ & $1.588$ & $1.533$ & $1.489$ & $1.444$ & $1.400$ & $1.331$ & $1.227$ & $1.067$ & $0.866$\\
$g(\lambda^{+})$ & $0.960$ & $0.929$ & $0.905$ & $0.877$ & $0.852$ & $0.807$ & $0.743$ & $0.646$ & $0.525$\\
\hline
\\ \multicolumn{10}{c} {$\lambda=1.50$}\\
 $T^{*}=1.5$\\
\cline{1-1}
$g(1^{+})$ & $1.952$ & $1.909$ & $1.884$ & $1.888$ & $1.989$ & $2.263$ & $2.783$ & $3.640$ & $4.962$\\
$g(\lambda^{-})$ & $1.832$ & $1.695$ & $1.570$ & $1.464$ & $1.382$ & $1.281$ & $1.150$ & $0.993$ & $0.850$\\
$g(\lambda^{+})$ & $0.941$ & $0.870$ & $0.804$ & $0.751$ & $0.709$ & $0.659$ & $0.590$ & $0.510$ & $0.437$\\
\\ $T^{*}=2.0$\\
\cline{1-1}
$g(1^{+})$ & $1.661$ & $1.660$ & $1.720$ & $1.820$ & $2.006$ & $2.336$ & $2.880$ & $3.740$ & $5.013$\\
$g(\lambda^{-})$ & $1.563$ & $1.478$ & $1.403$ & $1.338$ & $1.274$ & $1.185$ & $1.063$ & $0.922$ & $0.790$\\
$g(\lambda^{+})$ & $0.945$ & $0.894$ & $0.851$ & $0.814$ & $0.771$ & $0.719$ & $0.646$ & $0.558$ & $0.480$\\
\\ $T^{*}=3.0$\\
\cline{1-1}
$g(1^{+})$ & $1.444$ & $1.521$ & $1.628$ & $1.795$ & $2.046$ & $2.424$ & $2.998$ & $3.825$ & $5.025$\\
$g(\lambda^{-})$ & $1.351$ & $1.313$ & $1.268$ & $1.230$ & $1.171$ & $1.090$ & $0.982$ & $0.857$ & $0.728$\\
$g(\lambda^{+})$ & $0.968$ & $0.938$ & $0.909$ & $0.878$ & $0.839$ & $0.781$ & $0.704$ & $0.614$ & $0.521$\\
\hline
\\ \multicolumn{10}{c} {$\lambda=1.60$}\\
 $T^{*}=1.5$\\
\cline{1-1}
$g(1^{+})$ & $2.064$ & $2.026$ & $1.949$ & $1.894$ & $2.004$ & $2.339$ & $3.002$ & $3.996$ & $5.382$\\
$g(\lambda^{-})$ & $1.869$ & $1.696$ & $1.528$ & $1.387$ & $1.293$ & $1.200$ & $1.099$ & $1.015$ & $0.956$\\
$g(\lambda^{+})$ & $0.958$ & $0.872$ & $0.784$ & $0.713$ & $0.665$ & $0.618$ & $0.565$ & $0.521$ & $0.491$\\
\\ $T^{*}=2.0$\\
\cline{1-1}
$g(1^{+})$ & $1.655$ & $1.673$ & $1.702$ & $1.795$ & $2.007$ & $2.393$ & $3.037$ & $3.985$ & $5.313$\\
$g(\lambda^{-})$ & $1.542$ & $1.446$ & $1.350$ & $1.271$ & $1.199$ & $1.110$ & $1.015$ & $0.932$ & $0.872$\\
$g(\lambda^{+})$ & $0.940$ & $0.873$ & $0.819$ & $0.770$ & $0.726$ & $0.673$ & $0.615$ & $0.564$ & $0.530$\\
\\ $T^{*}=3.0$\\
\cline{1-1}
$g(1^{+})$ & $1.439$ & $1.498$ & $1.602$ & $1.762$ & $2.040$ & $2.463$ & $3.087$ & $3.987$ & $5.258$\\
$g(\lambda^{-})$ & $1.329$ & $1.269$ & $1.218$ & $1.165$ & $1.100$ & $1.021$ & $0.933$ & $0.850$ & $0.792$\\
$g(\lambda^{+})$ & $0.953$ & $0.912$ & $0.873$ & $0.834$ & $0.788$ & $0.732$ & $0.669$ & $0.611$ & $0.567$\\
\hline
\\ \multicolumn{10}{c} {$\lambda=1.70$}\\
 $T^{*}=2.0$\\
\cline{1-1}
$g(1^{+})$ & $1.716$ & $1.714$ & $1.752$ & $1.833$ & $2.077$ & $2.538$ & $3.250$ & $4.253$ & $5.611$\\
$g(\lambda^{-})$ & $1.551$ & $1.432$ & $1.322$ & $1.237$ & $1.165$ & $1.101$ & $1.045$ & $1.017$ & $1.022$\\
$g(\lambda^{+})$ & $0.944$ & $0.870$ & $0.802$ & $0.749$ & $0.706$ & $0.668$ & $0.634$ & $0.617$ & $0.619$\\
\\ $T^{*}=3.0$\\
\cline{1-1}
$g(1^{+})$ & $1.430$ & $1.494$ & $1.598$ & $1.779$ & $2.083$ & $2.549$ & $3.222$ & $4.160$ & $5.454$\\
$g(\lambda^{-})$ & $1.318$ & $1.251$ & $1.184$ & $1.131$ & $1.069$ & $1.009$ & $0.956$ & $0.928$ & $0.933$\\
$g(\lambda^{+})$ & $0.949$ & $0.893$ & $0.849$ & $0.810$ & $0.766$ & $0.725$ & $0.685$ & $0.664$ & $0.667$\\
\\ $T^{*}=5.0$\\
\cline{1-1}
$g(1^{+})$ & $1.299$ & $1.403$ & $1.554$ & $1.779$ & $2.114$ & $2.567$ & $3.220$ & $4.108$ & $5.351$\\
$g(\lambda^{-})$ & $1.177$ & $1.139$ & $1.097$ & $1.052$ & $1.000$ & $0.941$ & $0.893$ & $0.860$ & $0.867$\\
$g(\lambda^{+})$ & $0.964$ & $0.934$ & $0.898$ & $0.861$ & $0.818$ & $0.771$ & $0.731$ & $0.703$ & $0.709$\\
\hline
\\ \multicolumn{10}{c} {$\lambda=1.80$}\\
 $T^{*}=2.0$\\
\cline{1-1}
$g(1^{+})$ & $1.839$ & $1.916$ & $1.880$ & $1.932$ & $2.199$ & $2.708$ & $3.487$ & $4.508$ & $5.886$\\
$g(\lambda^{-})$ & $1.599$ & $1.468$ & $1.326$ & $1.233$ & $1.175$ & $1.144$ & $1.141$ & $1.183$ & $1.312$\\
$g(\lambda^{+})$ & $0.974$ & $0.892$ & $0.805$ & $0.745$ & $0.714$ & $0.694$ & $0.693$ & $0.716$ & $0.793$\\
\\ $T^{*}=3.0$\\
\cline{1-1}
$g(1^{+})$ & $1.449$ & $1.522$ & $1.629$ & $1.828$ & $2.162$ & $2.670$ & $3.392$ & $4.325$ & $5.627$\\
$g(\lambda^{-})$ & $1.320$ & $1.238$ & $1.175$ & $1.124$ & $1.079$ & $1.046$ & $1.043$ & $1.079$ & $1.195$\\
$g(\lambda^{+})$ & $0.942$ & $0.888$ & $0.840$ & $0.803$ & $0.773$ & $0.751$ & $0.747$ & $0.772$ & $0.853$\\
\\ $T^{*}=5.0$\\
\cline{1-1}
$g(1^{+})$ & $1.297$ & $1.411$ & $1.562$ & $1.809$ & $2.152$ & $2.646$ & $3.318$ & $4.214$ & $5.465$\\
$g(\lambda^{-})$ & $1.169$ & $1.126$ & $1.085$ & $1.042$ & $1.005$ & $0.978$ & $0.970$ & $1.003$ & $1.107$\\
$g(\lambda^{+})$ & $0.956$ & $0.921$ & $0.886$ & $0.854$ & $0.822$ & $0.800$ & $0.794$ & $0.821$ & $0.906$\\
\hline
\\ \multicolumn{10}{c} {$\lambda=1.90$}\\
 $T^{*}=3.0$\\
\cline{1-1}
$g(1^{+})$ & $1.492$ & $1.570$ & $1.692$ & $1.908$ & $2.274$ & $2.815$ & $3.533$ & $4.450$ & $5.534$\\
$g(\lambda^{-})$ & $1.323$ & $1.247$ & $1.182$ & $1.142$ & $1.123$ & $1.132$ & $1.174$ & $1.265$ & $1.403$\\
$g(\lambda^{+})$ & $0.949$ & $0.895$ & $0.848$ & $0.818$ & $0.805$ & $0.811$ & $0.841$ & $0.908$ & $1.008$\\
\\ $T^{*}=5.0$\\
\cline{1-1}
$g(1^{+})$ & $1.309$ & $1.416$ & $1.587$ & $1.847$ & $2.222$ & $2.733$ & $3.404$ & $4.268$ & $5.353$\\
$g(\lambda^{-})$ & $1.170$ & $1.123$ & $1.087$ & $1.058$ & $1.044$ & $1.049$ & $1.089$ & $1.174$ & $1.305$\\
$g(\lambda^{+})$ & $0.958$ & $0.920$ & $0.889$ & $0.867$ & $0.855$ & $0.859$ & $0.891$ & $0.961$ & $1.072$\\
\hline
\\ \multicolumn{10}{c} {$\lambda=2.00$}\\
 $T^{*}=3.0$\\
\cline{1-1}
$g(1^{+})$ & $1.564$ & $1.699$ & $1.824$ & $2.032$ & $2.424$ & $2.973$ & $3.669$ & $4.493$ & $5.499$\\
$g(\lambda^{-})$ & $1.346$ & $1.282$ & $1.217$ & $1.192$ & $1.199$ & $1.245$ & $1.322$ & $1.425$ & $1.528$\\
$g(\lambda^{+})$ & $0.968$ & $0.918$ & $0.873$ & $0.853$ & $0.861$ & $0.892$ & $0.949$ & $1.024$ & $1.100$\\
\\ $T^{*}=5.0$\\
\cline{1-1}
$g(1^{+})$ & $1.325$ & $1.450$ & $1.633$ & $1.907$ & $2.300$ & $2.814$ & $3.471$ & $4.304$ & $5.354$\\
$g(\lambda^{-})$ & $1.178$ & $1.131$ & $1.105$ & $1.098$ & $1.109$ & $1.152$ & $1.225$ & $1.326$ & $1.428$\\
$g(\lambda^{+})$ & $0.960$ & $0.926$ & $0.905$ & $0.899$ & $0.910$ & $0.944$ & $1.005$ & $1.089$ & $1.175$\\
\label{Table g(1+)-g(lambda+-)}
\end{longtable}
\endgroup

\section{Results and discussion\label{sec4}}

 The objective is to determine the limits of applicability of the {TL
and  YS} theories. We are mainly interested in the domain of
\textit{moderate} temperatures. By that we mean temperatures within
the range $T_c^*(\lambda)\lesssim T^*\lesssim 3T_c^*(\lambda)$,
where $T_c^*(\lambda)$ denotes the critical temperature of the SW
fluid with range $\lambda$. This critical temperature has been
measured in computer simulations for several
ranges.\cite{VMRJM:92,M:97,BV:98-a,OP:99,EH:99,OFP:01,LAB:96} Table
I of Ref.\ \onlinecite{RK:02} gives a rather extensive compilation
of data. A simple analytical estimate for $T_c^*(\lambda)$ was
derived in Ref.\ \onlinecite{AS:01}:
\beq
T_c^*(\lambda)=\frac{1}{\ln\left[1+\frac{3+\lambda+2\sqrt{2\lambda}}{\lambda(\lambda-1)(9-2\lambda+\lambda^2)}\right]}.
\label{4.1}
\eeq

\begin{figure}[tbp]
\includegraphics[width=0.9\columnwidth]{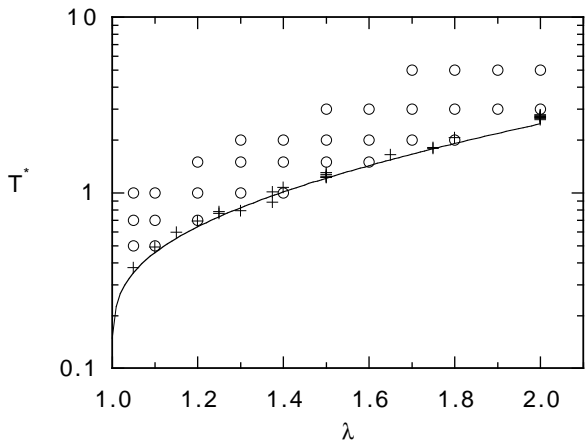}
\caption{The open circles represent the values of the reduced
temperature $T^*$ we have considered for each value of the range
$\lambda$. The crosses are simulation data for the critical
temperature $T_c^*(\lambda)$ (see Ref.\ \protect\onlinecite{RK:02})
and the solid line is the theoretical estimate (\protect\ref{4.1}).
Note the logarithmic scale of the vertical axis.}
\label{Fig_Tc}
\end{figure}
Figure \ref{Fig_Tc} is a $T^*$-$\lambda$ plot where the open circles
represent the three  temperatures (two in the cases $\lambda=1.9$
and $\lambda=2$) we have considered in the simulations for each
value of $\lambda$. The simulation data of
$T_c^*(\lambda)$,\cite{RK:02} as well as the theoretical estimate
(\ref{4.1}) are also shown. For each value of $\lambda$ we have
typically considered three temperatures: $T_1^*\gtrsim T_c^*$,
$T_2^*\approx 1.5 T_1^*$, and $T_3^*\approx 2 T_1^*$, {so that
$T_3^*\gtrsim 3T_c^*(\lambda)$.}

\begin{figure}[tbp]
\includegraphics[width=0.9\columnwidth]{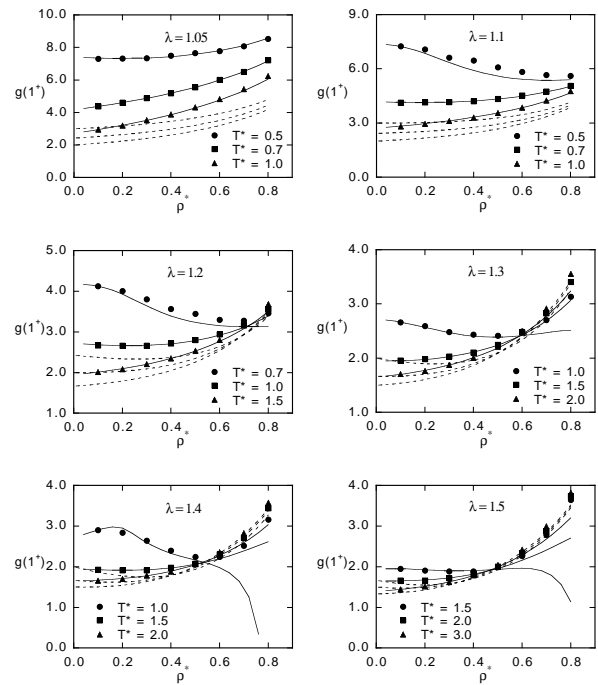}
\caption{Comparison of the contact values $g(1^+)$ of the r.d.f.
obtained from the TL (dashed lines) and YS (solid lines) theories
with Monte Carlo data  as  functions of the reduced density
$\rho^{*}$ for different temperatures and well widths
$\lambda=1.05$--$1.5$.}
\label{Fig_g1}
\end{figure}
\begin{figure}[tbp]
\includegraphics[width=0.9\columnwidth]{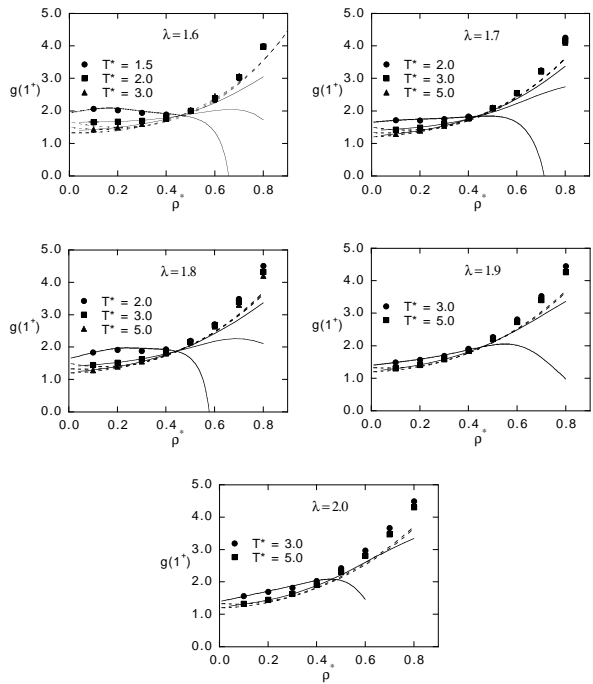}
\caption{Same as in Fig.\ \protect\ref{Fig_g1}, but for
$\lambda=1.6$--$2$.}
\label{Fig_g1bis}
\end{figure}

Before comparing the simulation data of the full r.d.f. with the
theoretical predictions, it is worth focusing on the contact value
$g(1^+)$. Figures \ref{Fig_g1} and \ref{Fig_g1bis} show $g(1^+)$ as
a function of the reduced density for the temperatures represented
in Fig.\ \ref{Fig_Tc}, as obtained from our MC simulations, as well
as from the TL and YS theories [cf.\ Eqs.\ (\ref{A4bis}),
(\ref{A5bis}), and (\ref{B0})]. We observe that for $\lambda=1.05$
and $\lambda=1.1$ the non-perturbative YS model presents a very good
agreement with the simulation data for the three temperatures
considered, whereas the TL perturbation  theory is rather poor,
especially for low temperatures. For $\lambda\geq 1.2$, however, the
YS model behaves well for small and moderate densities but starts to
fail in the high-density domain, especially for the lowest
temperature, the failure being more dramatic as the well width
increases. Interestingly, the TL theory becomes more accurate
precisely in that high-density region where the YS model is less
reliable.  Thus, for a given range $\lambda$, there exists a certain
threshold density $\rho_0^*(\lambda)$ such that the YS model is
accurate for $\rho^*\lesssim \rho_0^*(\lambda)$ and inaccurate for
$\rho^*\gtrsim \rho_0^*(\lambda)$, while the opposite situation
occurs in the case of the TL theory. Of course, this qualitative
description applies for the range of ``moderate'' temperatures
defined above, since {the results obtained from} both theories tend
to coincide as the temperature increases.

According to Figs.\ \ref{Fig_g1} and \ref{Fig_g1bis}, the location
of $\rho_0^*(\lambda)$ roughly coincides with the region where
either the isotherms cross  (for $\lambda \leq 1.7$) or have the
least separation (for $1.8 \leq \lambda \leq 2.0$). This means that
the simulation data of $g(1^+)$ in the region
$\rho^*\approx\rho_0^*(\lambda)$ are practically insensitive to the
temperature, so they are close to its hard-sphere value $g_0(1^+)$.
For larger densities, $\rho^*\gtrsim \rho_0^*(\lambda)$, the
simulation data show that the influence of temperature is small and
hence the TL perturbation theory becomes accurate in that domain. On
the other hand, for $\rho^*\lesssim\rho_0^*(\lambda)$ the MC values
of $g(1^{+})$ are strongly sensitive to temperature, as expected
from the fact that at zero density {$g(1^+)=e^{1/T^*}$, while
perturbation theories give $g(1^+)=1+1/T^*$. The strong deviation of
the non-perturbative YS theory from the MC data in the density
region $\rho^*\gtrsim \rho_0^*(\lambda)$, especially for
$\lambda\geq 1.4$, is in part due to the fact that in the YS model
the parameter $A$ in Eq.\ (\ref{eq:F(t)}) is assumed to be
independent of density and so it is assigned its zero-density value
$A=e^{1/T^*}-1$. A better agreement is expected if $A$ is allowed to
depend on density, but this would imply either to impose an extra
consistency condition (for instance, continuity of the first
derivative of the cavity function) or to apply an empirical fit,
what is outside the original spirit of the YS model. A second reason
has to do with the construction of the YS model as an extension of
Baxter's solution of the PY integral equation for sticky hard
spheres, so that in principle it is intended to be a model for
narrow wells.

\begin{figure}[tbp]
\includegraphics[width=0.9\columnwidth]{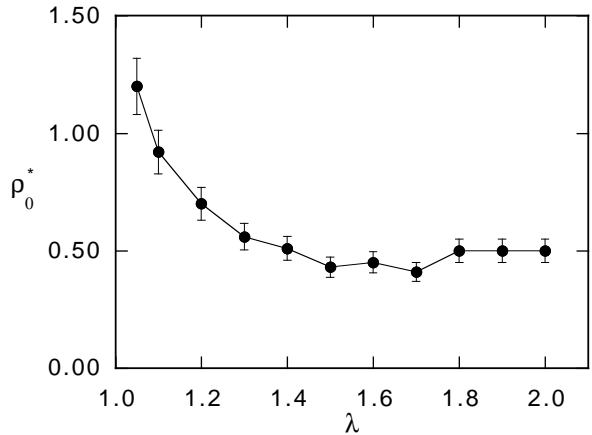}
\caption{Plot of the threshold density $\rho_0^*(\lambda)$ as a
function of the potential range $\lambda$. For each value of
$\lambda$, $\rho_0^*$ is defined as the density around which the MC
contact values $g(1^+)$ are practically insensitive to the
temperature. Below (above) the curve, the YS (TL) theory can be
considered as reliable. The line is a guide to the eye.}
\label{Fig_rho0}
\end{figure}

A plot of $\rho_0^*(\lambda)$ is presented in Fig.\ \ref{Fig_rho0}.
It can be interpreted as a sort of ``phase'' diagram  in which the
curve separates the respective regions where the TL and YS theories
are reliable for moderate temperatures in the interval
$T_c^*(\lambda)\lesssim T^*\lesssim 3 T_c^*(\lambda)$. We observe
that as the range $\lambda$ decreases, the YS region tends to span
the whole fluid density domain. In addition, $\rho_0^*$ presents a
minimum $\rho_0^*\approx 0.4$ at $\lambda\approx 1.7$, so the YS
theory {does} a fairly good job if $\rho^*\lesssim 0.4$, even for
wide potentials.

\begin{figure}[tbp]
\includegraphics[width=0.9\columnwidth]{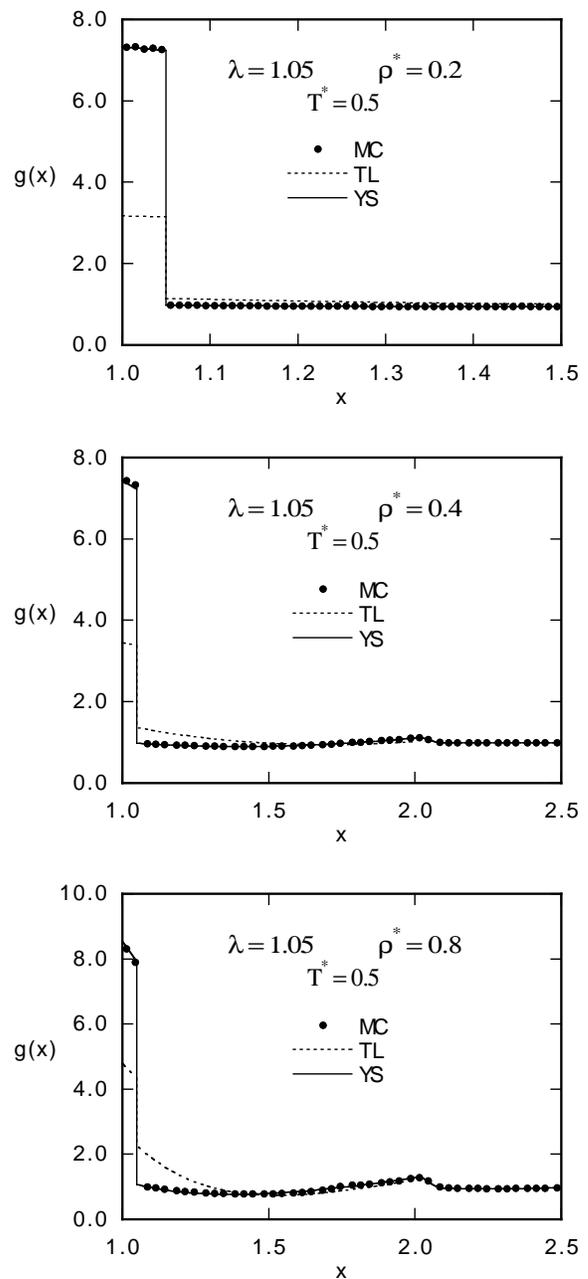}
\caption{Comparison of the r.d.f. obtained from the TL (dotted
lines) and YS (solid lines) theories with Monte Carlo data (circles)
for $\lambda=1.05$ and $T^{*}=0.5$. {Note that the TL curves are
interrupted after $x=2$.}}
\label{Fig_gx_105}
\end{figure}
\begin{figure}[tbp]
\includegraphics[width=0.9\columnwidth]{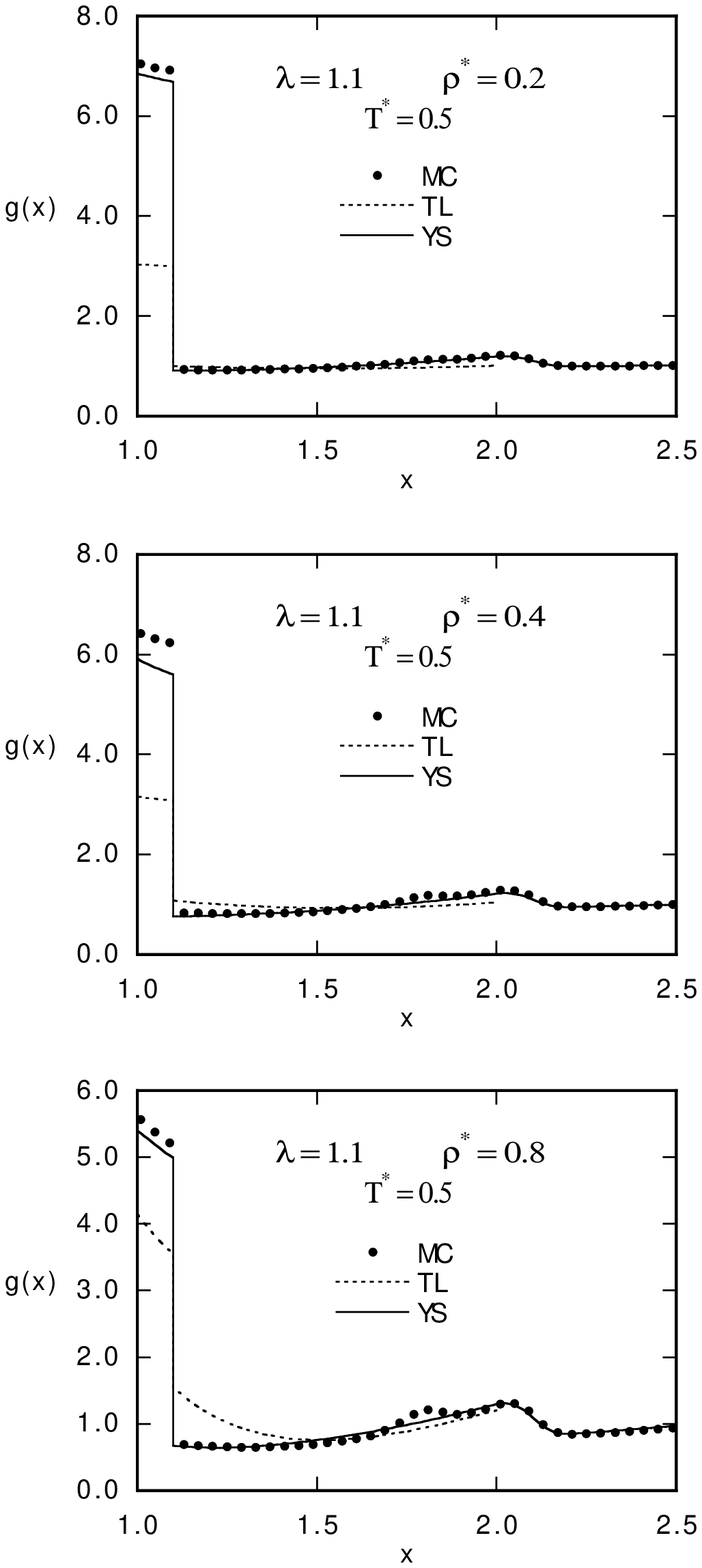}
\caption{Same as in Fig.\ \protect\ref{Fig_gx_105}, but for
$\lambda=1.1$ and $T^{*}=0.5$.}
\label{Fig_gx_11}
\end{figure}
\begin{figure}[tbp]
\includegraphics[width=0.9\columnwidth]{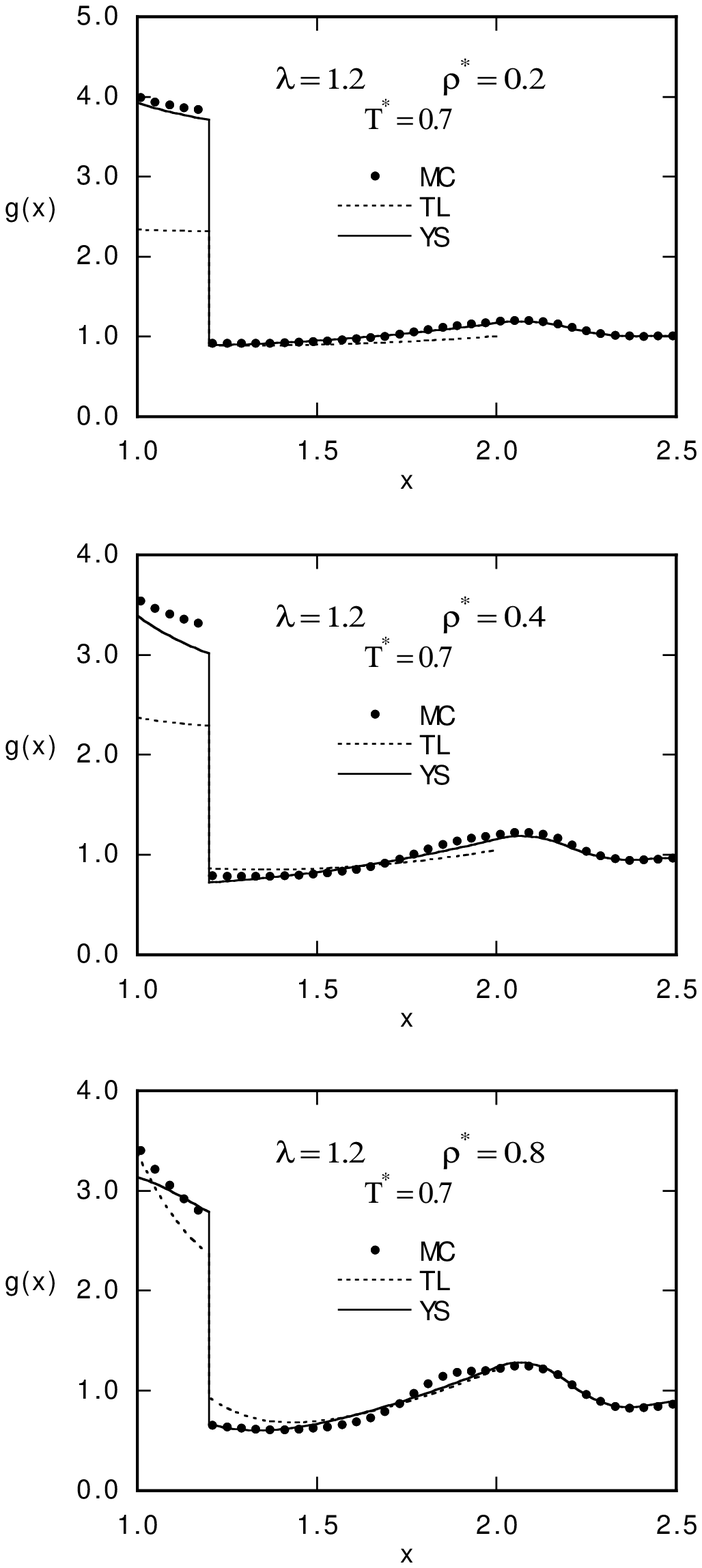}
\caption{Same as in Fig.\ \protect\ref{Fig_gx_105}, but for
$\lambda=1.2$ and $T^{*}=0.7$.}
\label{Fig_gx_12}
\end{figure}
\begin{figure}[tbp]
\includegraphics[width=0.9\columnwidth]{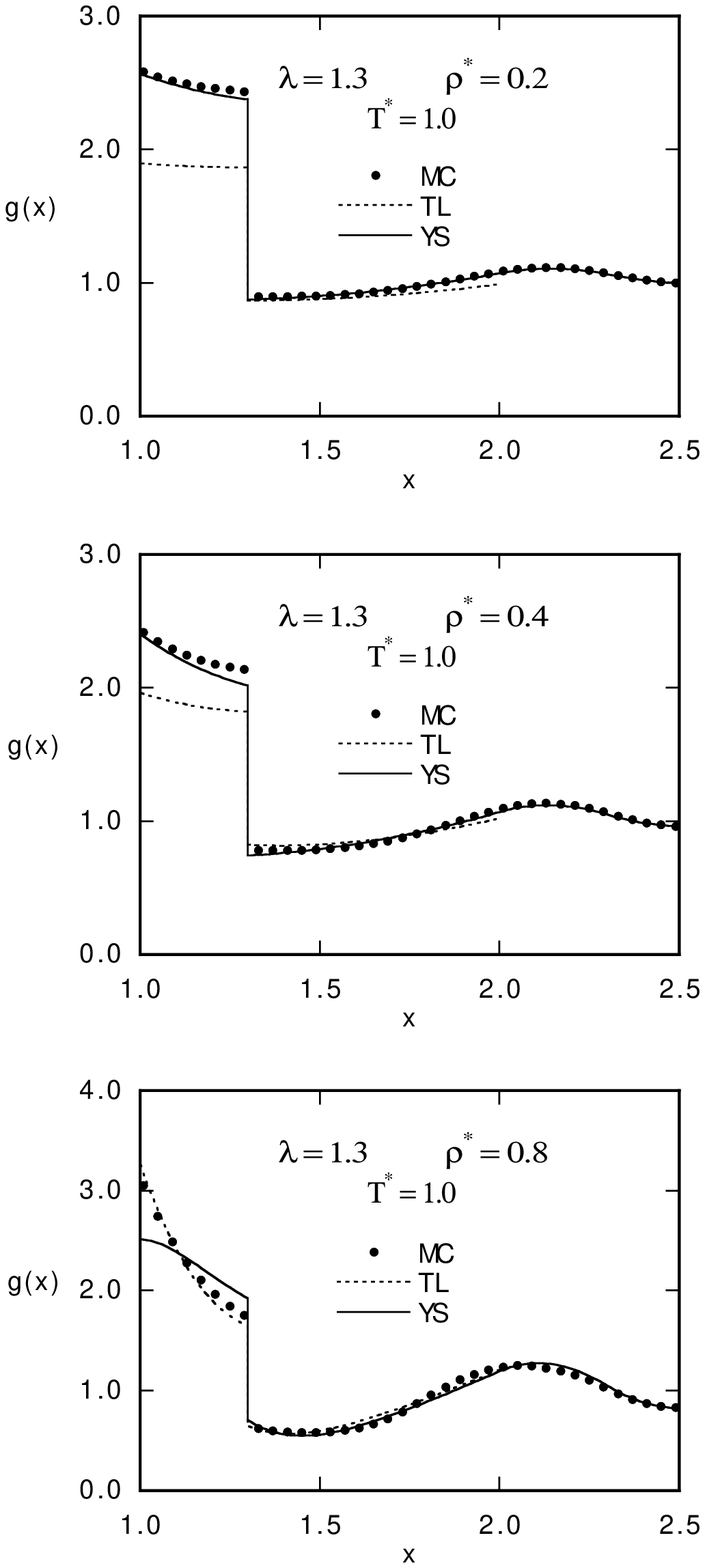}
\caption{Same as in Fig.\ \protect\ref{Fig_gx_105}, but for
$\lambda=1.3$ and $T^{*}=1.0$.}
\label{Fig_gx_13}
\end{figure}
\begin{figure}[tbp]
\includegraphics[width=0.9\columnwidth]{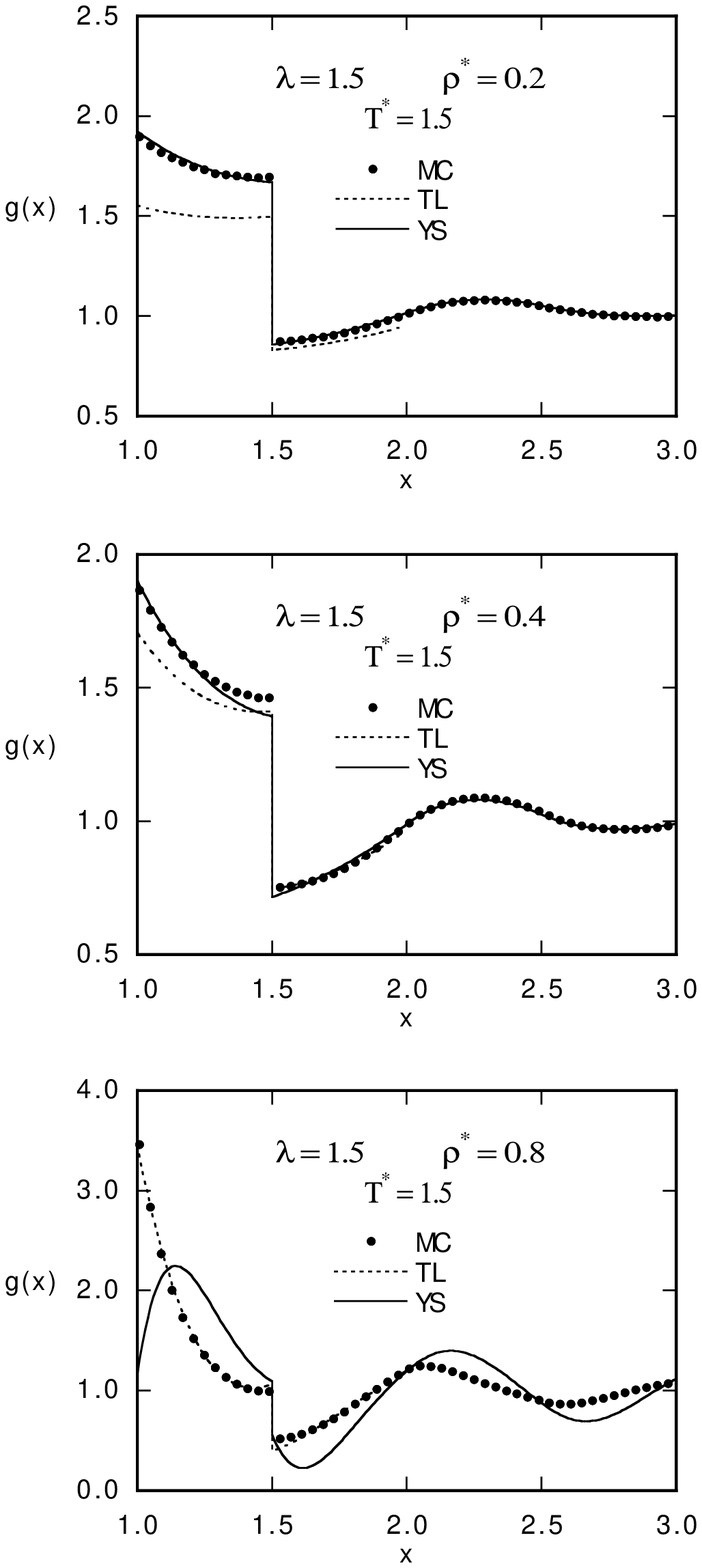}
\caption{Same as in Fig.\ \protect\ref{Fig_gx_105}, but  for
$\lambda=1.5$ and $T^{*}=1.5$.}
\label{Fig_gx_15}
\end{figure}
\begin{figure}[tbp]
\includegraphics[width=0.9\columnwidth]{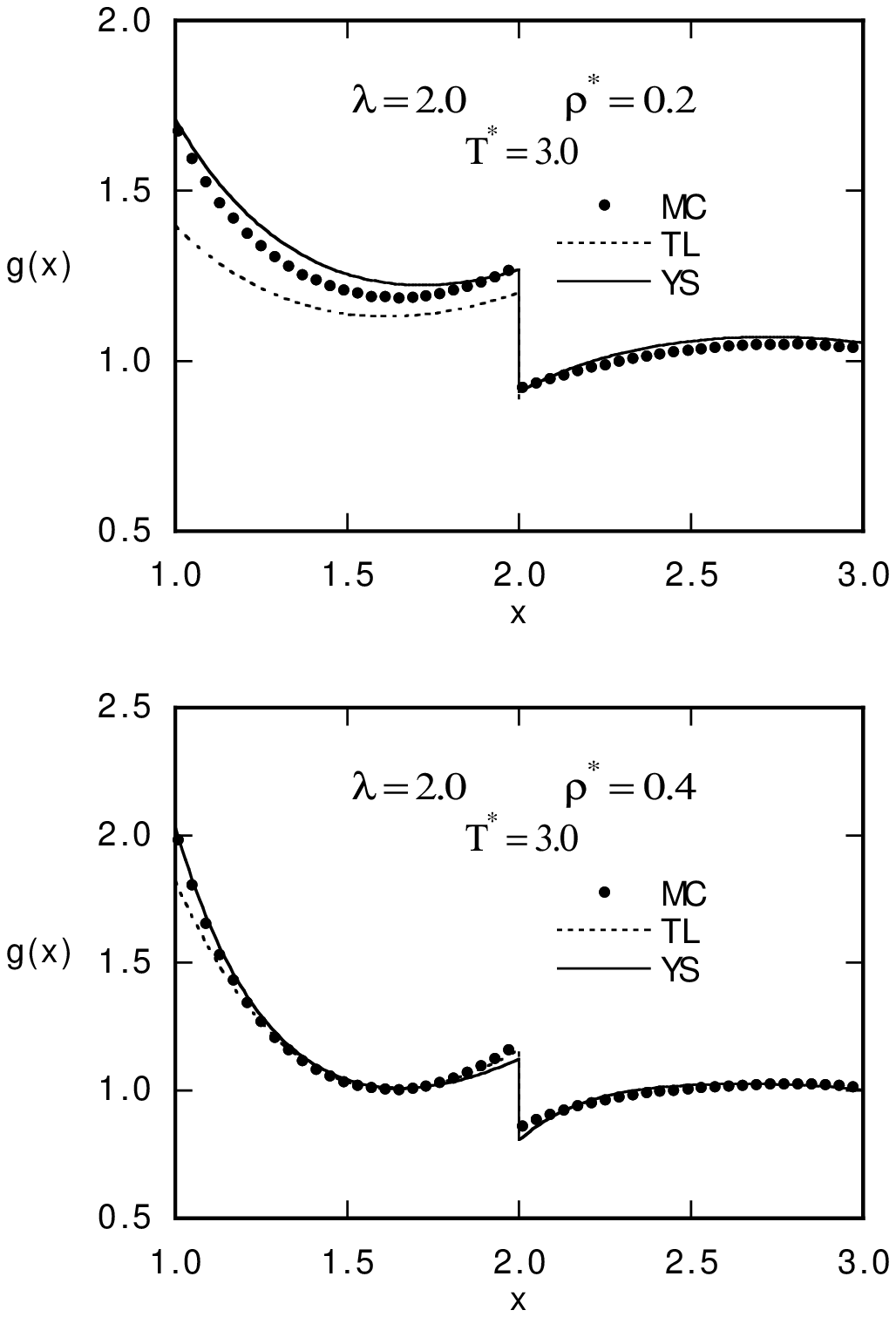}
\caption{Same as in Fig.\ \protect\ref{Fig_gx_105}, but  for
$\lambda=2.0$ and $T^{*}=3.0$.}
\label{Fig_gx_20}
\end{figure}

Once we have analyzed the performances of the TL and YS theories in
connection with the contact value $g(1^+)$, let us proceed to
investigate the r.d.f. $g(x)$ itself. The results are presented in
Figs.\ \ref{Fig_gx_105}--\ref{Fig_gx_20}. Since in this paper we are
mainly interested in short-ranged SW potentials, we have paid
special attention to the ranges $1.05\leq \lambda\leq 1.3$ (Figs.\
\ref{Fig_gx_105}--\ref{Fig_gx_13}). As representative examples of a
moderate and of a wide range we have considered $\lambda=1.5$ (Fig.\
\ref{Fig_gx_15}) and $\lambda=2.0$ (Fig.\ \ref{Fig_gx_20}),
respectively. For each value of $\lambda$ we have restricted
ourselves to the lowest temperature represented in Fig.\
\ref{Fig_Tc} and to the densities $\rho^*=0.2$, $0.4$, and $0.8$
(except in the case $\lambda=2.0$, where $\rho^*=0.8$ has not been
considered because the YS model fails to have a solution in that
case). {In agreement with} the analysis of Figs.\ \ref{Fig_g1} and
\ref{Fig_g1bis},  one can see that the YS theory works  well for
small potential widths ($\lambda \le 1.2$) for the whole density
range. For larger potential widths, the performance of the theory is
still fair at low ($\rho^{*}= 0.2$) and even moderate ($\rho^{*}=
0.4$) densities. However, the YS theory fails, and even {can become}
entirely unphysical, at high densities ($\rho^{*}=0.8
>\rho_0^*$). Of course, at  temperatures higher than those of Figs.\
\ref{Fig_gx_105}--\ref{Fig_gx_20} the performance of the theory at
high densities improves (not shown).

By contrast, the TL theory presents the  {opposite behavior} to that
of the YS theory, since its accuracy increases as the density and
the potential width grow. Of course, it also improves if the
temperature increases, as expected from a perturbation theory.
According to Figs.\ \ref{Fig_gx_105}--\ref{Fig_gx_20}, the TL theory
does a better job than the YS model at $\rho^*=0.8$ for $\lambda\geq
1.3$, in agreement with the ``phase'' diagram of Fig.\
\ref{Fig_rho0}.

\section{Conclusions\label{sec5}}
 In this paper we have presented extensive Monte Carlo simulations for the structural properties of
square-well  fluids with ranges $\lambda$, reduced densities
$\rho^*$, and reduced temperatures $T^*$  in the intervals $1.05\leq
\lambda\leq 2$, $0.1\leq\rho^*\leq 0.8$ and $T_c^*(\lambda)\lesssim
T^*\lesssim 3T_c^*(\lambda)$, respectively. The MC data have been
used to assess the accuracy of two theories that provide explicit
expressions of the r.d.f. in Laplace space, the TL perturbation
theory\cite{TL:94-a,TL:94-b} and the non-perturbative YS
model.\cite{BYS:94}

The results show that both theories complement each other, as the YS
theory works well where the TL theory fails and vice versa. More
specifically, the YS theory exhibits a good agreement with the MC
data at any fluid density if the potential well is sufficiently
narrow (say $\lambda\leq 1.2$), as well as for any width if the
density is small enough (say $\rho^*\leq 0.4$). This can be further
refined by {noticing} that the YS theory works well if
$\rho^*\lesssim \rho_0^*(\lambda)$, where $\rho_0^*(\lambda)$ is the
density around which the simulation data for the contact value
$g(1^+)$ show the least influence on temperature. On the other hand,
for $\rho^*>\rho_0^*(\lambda)$ the YS theory rapidly deteriorates,
especially for temperatures near the critical one, while the TL
theory becomes very accurate.

The complementarity between the TL and YS theories is interesting
because they present some formal similarities in their formulation
and are (practically) equally easy to implement (see Appendices
\ref{appA} and \ref{appB}). The latter theory, however, has some
advantages over the former one. First, the YS theory is especially
useful for describing colloidal dispersions modeled as short-ranged
SW fluids. Second, it provides a simple analytical expression for
the second shell ($2 \le x \le 3$) of the r.d.f., whereas this is
not the case for the TL theory.\cite{TL:94-b} Last, it seems
feasible to improve the performance of the YS theory at high
densities by imposing additional constraints to the Laplace
transform of the r.d.f. to determine the parameter $A$ in Eq.\
(\ref{eq:F(t)}) as a function of density. Instead, in order to
improve the TL theory it would be necessary to obtain higher order
terms in the expansion of the r.d.f. of the SW fluid in power series
of the inverse of the reduced temperature $T^{*}$, and this seems to
be too complicated at present.

\acknowledgments

The present work has been partially supported by the Spanish Direcci\'{o}n
General de
Investigaci\'{o}n (DGI) under grants No.\  BFM2003-001903 (J.L. and J.R.S) and  No.\
FIS2004-01399 (S.B.Y. and A.S.).

\appendix

\section{Explicit expressions in the Tang--Lu theory\label{appA}}
 Let us define the Laplace transform $G(s)$ of
$xg(x)$:
\beq
G(t)=\int_1^\infty dx\, e^{-tx} xg(x).
\label{A1}
\eeq
The contact value $g(1^+)$ is given from $G(s)$ as
\beq
g(1^+)=\lim_{t\to \infty} te^t G(t).
\label{A1bis}
\eeq

The exact solution of the PY equation for hard
spheres\cite{W:63,T:63} reads
\beq
G_0(t)=t\frac{L(t)e^{-t}}{S(t)+12\eta L(t)e^{-t}},
\label{A2}
\eeq
where $\eta=\frac{\pi}{6}\rho\sigma^3$ is the packing fraction and
\beq
L(t)=1+2\eta+(1+\eta/2)t,
\label{A3}
\eeq
\beq
S(t)=-12\eta(1+2\eta)+18\eta^2 t+6\eta(1-\eta)t^2+(1-\eta)^2t^3.
\label{A4}
\eeq
The corresponding contact value is
\beq
g_0(1^+)=\frac{1+\eta/2}{(1-\eta)^2}.
\label{A4bis}
\eeq

Equation (\ref{A2})  provides the zeroth-order term in the TL
perturbation theory. The first-order term is\cite{TL:94-a,TL:94-b}
\beqa
G_1(t)&=&-\frac{(1-\eta)^4e^{-t}}{Q_0^2(t)}\left\{\frac{t^4(1+\lambda
t)}{S^2(-t)}e^{-(\lambda-1)t} \right.  \nn & &
 -
\sum_{i=1}^3\frac{t_i^3}{(t+t_i)S_1^2(t_i)}\left[\frac{t_i(1-\lambda
t_i)}{t+t_i}+t_i(1-\lambda t_i)\right.\nn
&&\left.\times\frac{S_2(t_i)}{S_1(t_i)}
-4+(1+4\lambda)t_i+\lambda(\lambda-1)t_i^2\right]\nn &&\left.\times
e^{(\lambda-1)t_i}\right\},
\label{A5}
\eeqa
where $S_1(t)\equiv S'(t)$, $S_2(t)\equiv S''(t)$, the primes
denoting derivatives with respect to $t$, and
\beq
Q_0(t)\equiv\frac{S(t)+12\eta L(t)e^{-t}}{(1-\eta)^2t^3}.
\label{A6}
\eeq
In Eq.\ (\ref{A5}) the summation extends over the three zeroes of
$S(t)$, denoted by $t_i$. The contact value $g_1(1^+)$ is
\beqa
g_1(1^+)&=&(1-\eta)^4
\sum_{i=1}^3\frac{t_i^3}{S_1^2(t_i)}\left[t_i(1-\lambda
t_i)\frac{S_2(t_i)}{S_1(t_i)} -4\right. \nn
&&\left.+(1+4\lambda)t_i+\lambda(\lambda-1)t_i^2\right]
e^{(\lambda-1)t_i}
\label{A5bis}
\eeqa

By analytical inversion of $G_1(t)$ one can get explicit expressions
for $g_1(x)$ inside the shells $n\leq x \leq n+1$, which become
increasingly more complicated as $n$ grows. The expression for the
first shell $1\leq x\leq 2$ can be found in Ref.\
\onlinecite{TL:94-b}.

\section{Explicit expressions in the Yuste--Santos model\label{appB}}

 By imposing the exact condition $G(t)-t^{-2}\sim t$ for small $t$, where $G(t)$ is defined by Eq.\ (\ref{A1}), one can
express  the parameters $K_1$, $S_1$, $S_2$, and $S_3$ appearing in
Eq.\ (\ref{eq:F(t)}) as linear functions of $A$ and
$K_2$:\cite{BYS:94,AS:01}
\beqa
\label{c5}
K_1&=&\frac{1}{1+2\eta}\left[1+\frac{\eta}{2}+2\eta(\lambda^3-1)
K_2\right.\nn
 &&\left.-\frac{\eta}{2}(\lambda^4-4\lambda+3)
{A}\right]+K_2-A(\lambda-1),
\eeqa
\begin{equation}
\label{c6}
S_1=\frac{\eta}{1+2\eta}\left[-\frac{3}{2}+2(\lambda^3-1)
K_2-
\frac{1}{2}(\lambda^4-4\lambda+3)
{A}\right],
\end{equation}
\beqa
\label{c7}
S_2&=&\frac{1}{2(1+2\eta)}\left\{-1+\eta+2\left[\lambda-1-2\eta\lambda(\lambda^2-1)\right]K_2
\right.\nn
&&\left.-\left[(\lambda-1)^2-\eta(\lambda^2-1)^2\right]{A}\right\},
\eeqa
\beqa
\label{c7bis}
S_3&=&\frac{1}{1+2\eta}\left\{-\frac{(1-\eta)^2}{12\eta}-
\left[\frac{1}{2}(\lambda^2-1)-\eta\lambda^2(\lambda-1)\right]K_2\right.\nn
&&\left.
+\frac{1}{12}[4+2\lambda-\eta(3\lambda^2+2\lambda+1)](\lambda-1)^2{A}\right\}.
\eeqa
{}From Eq.\ (\ref{A1bis}), we have
\beq
g(1^+)=\frac{K_1}{12\eta S_3}.
\label{B0}
\eeq

By application of the Heaviside expansion theorem, the inverse
Laplace transform of $tF(t)$ reads
\beq
f_1(x)=f_{10}(x)\Theta(x)+f_{11}(x+1-\lambda)\Theta(x+1-\lambda),
\label{B1}
\eeq
where
\beq
f_{1k}(x)=-\frac{1}{12\eta}\sum_{i=1}^3
\frac{C_{1k}(t_i)}{S'(t_i)}t_i e^{t_i x}.
\label{B2}
\eeq
Here, $t_i$ are the three distinct roots of $S(t)\equiv 1+S_1
t+S_2 t^2+S_3 t^3$ [not to be confused with the polynomial
(\ref{A4})] and
\beq
C_{10}(t)\equiv 1+A+K_1 t,\quad C_{11}(t)\equiv -(A+K_2 t).
\label{B3}
\eeq

Analogously, the inverse Laplace transform of $t[F(t)]^2$ is
\beqa
f_2(x)&=&f_{20}(x)\Theta(x)+f_{21}(x+1-\lambda)\Theta(x+1-\lambda)\nn
&&+f_{22}(x+2-2\lambda)\Theta(x+2-2\lambda),
\label{B4}
\eeqa
where
\beqa
f_{2k}(x)&=&\frac{1}{(12\eta)^2}\sum_{i=1}^3 \left[x
C_{2k}(t_i)+C_{2k}'(t_i)\right.\nn
&&\left.-C_{2k}(t_i)\frac{S''(t_i)}{S'(t_i)}\right]\frac{e^{t_i
x}}{[S'(t_i)]^2},
\label{B5}
\eeqa
where we have called
\beq
\begin{aligned}
C_{20}(t)&\equiv t[C_{10}(t)]^2,\quad C_{21}(t)\equiv 2t
C_{10}(t)C_{11}(t), \\ C_{22}(t)&\equiv t[C_{11}(t)]^2.
\end{aligned}
\label{B6}
\eeq

Insertion of Eqs.\ (\ref{B1}) and (\ref{B4}) into Eq.\
(\ref{eq:g(r)-SW}) gives the r.d.f. $g(x)$ in the interval $1\leq
x\leq 3$. Note that the contribution $f_{22}(x)$ is needed inside
that interval only if $\lambda<\frac{3}{2}$. For $x>3$ the
evaluation of $f_3(x), f_4(x), \ldots$ is required. Alternatively,
one can make use of the efficient method discussed by  Abate and
Whitt\cite{AW:92} to invert numerically Laplace transforms.

To close the model, we need to determine the parameters $A$ and
$K_2$. The former is assigned its zero-density limit value, namely
$A=e^{1/T^*}-1$.\cite{BYS:94} To determine $K_2$ we impose the
continuity condition of the cavity function at $x=\lambda$, what
implies
\beq
g(\lambda^-)=e^{1/T^*}g(\lambda^-).
\label{B8}
\eeq
 This yields
\beq
\left(1-e^{-1/T^*}\right)f_{10}(\lambda-1)=-f_{11}(0)=-\frac{K_2}{12\eta S_3}.
\label{B7}
\eeq
Since the roots $t_i$ depend on $K_2$ through the coefficients
$S_1$, $S_2$, and $S_3$, Eq.\ (\ref{B7}) is a {transcendent}
equation for $K_2$ that needs to be solved numerically. Acedo and
Santos have recently proposed a simplified version of the YS model
whereby the exact condition (\ref{B8}) is replaced by  a simpler one
that allows  $K_2$ to be obtained analytically.\cite{AS:01} This is
especially useful {for determining} the thermodynamic
properties.\cite{LSAS:03,AS:01} In this paper, however, since we are
interested in the structural properties, we enforce condition
(\ref{B8}) and determine $K_2$ from Eq.\ (\ref{B7}).


\begin{thebibliography}{99}
\bibitem{note}
One can check that the simulation values of {the ratio
$g(\lambda^-)/g(\lambda^+)$} deviate from the exact relationship
$g(\lambda^-)/g(\lambda^+)=e^{1/T^*}$ less than 0.6\% in all the
cases and less than 0.2\% in most of the cases.
\end{thebibliography}
\end{document}